\documentclass[a4paper,11pt,
colorlinks=true,urlcolor=blue,anchorcolor=blue,citecolor=blue,filecolor=blue,linkcolor=blue,menucolor=blue,linktocpage=true,pdfa=true,hidelinks]{article}
\usepackage{jinstpub} 
\usepackage{orcidlink}
\makeatletter
\let\frontmatter@title@above\relax
\makeatother

\newcommand{\sisero}[0]{\mbox{SiSeRO}}

\usepackage{soul}
\usepackage[utf8]{inputenc}
\usepackage[T1]{fontenc}

\begin{document}

\title{Automating Sensor Characterization with Bayesian Optimization}

\author[1,2]{J. Cuevas-Zepeda\,\orcidlink{0000-0002-2358-7049}}
\author[3]{C. Chavez\,\orcidlink{0000-0002-7853-6900}}

\author[2,3,4]{J. Estrada} 
\author[5]{J. Noonan} 
\author[2,3,6]{B. D. Nord\,\orcidlink{0000-0001-6706-8972}} 
\author[3,6]{N. Saffold\,\orcidlink{0000-0001-6358-9228}} 
\author[7]{M. Sofo-Haro\,\orcidlink{orcid=0000-0001-9397-2922}} 
\author[2]{R. Spinola e Castro\,\orcidlink{0009-0006-8010-9410}} 
\author[3]{S. Trivedi\,\orcidlink{0000-0003-1237-4301}}

\affiliation[1]{Kavli Institute for Cosmological Physics, University of Chicago, Chicago, IL 60637, USA}
\affiliation[2]{Department of Astronomy and Astrophysics, University of Chicago, Chicago, IL 60637, USA}
\affiliation[3]{Fermi National Accelerator Laboratory, Batavia, IL 60510, USA}
\affiliation[4]{Instrumentation Division, Brookhaven National Laboratory, Upton, NY 11973, USA}
\affiliation[5]{Department of Physics, University of Chicago, Chicago, IL 60637, USA}
\affiliation[6]{Kavli Institute for Cosmological Physics, University of Chicago, Chicago, IL 60637, USA}
\affiliation[7]{Comisión Nacional de Energía Atómica (CNEA) y Consejo Nacional de Investigaciones Científicas y Técnicas (CONICET), Universidad Nacional de Córdoba, Córdoba 5000, Argentina}

\emailAdd{juliancz@uchicago.edu}

\abstract{
The development of novel instrumentation requires an iterative cycle with three stages: design, prototyping, and testing. 
Recent advancements in simulation and nanofabrication techniques have significantly accelerated the design and prototyping phases.
Nonetheless, detector characterization continues to be a major bottleneck in device development.
During the testing phase, a significant time investment is required to characterize the device in different operating conditions and find optimal operating parameters. 
The total effort spent on characterization and parameter optimization can occupy a year or more of an expert’s time. 
In this work, we present a novel technique for automated sensor characterization that aims to accelerate the testing stage of the development cycle.
This technique leverages closed-loop Bayesian optimization (BO), using real-time measurements to guide parameter selection and identify optimal operating states.
We demonstrate the method with a novel low-noise CCD, showing that the machine learning-driven tool can efficiently characterize and optimize operation of the sensor in a couple of days without supervision of a device expert. 

}

\maketitle
\flushbottom

\section{Introduction}
\label{sec:intro}

Most photonic sensors are solid-state devices that measure radiation-induced signals --- charge and/or phonons --- with on-chip active components. 
Photonic sensors are the primary instrument for data acquisition in many contexts, including high-energy physics, cosmology, medicine, material science, and various industrial applications.
Further improvements in scientific and technical performance require surpassing current sensor limitations and streamlining design, fabrication, and characterization workflows.

There is a standard development and implementation cycle for sensors.
First, designers use simulations, such as Technology Computer-Aided Design (TCAD) \citep{Selberherr1984,Sentaurus}, to guide sensor design, modeling electrostatics, charge transfer, and noise behavior.
Second, the sensor is fabricated at a nanofabrication facility, a process that typically takes several weeks depending on device complexity.
Third, sensor characterization is performed on custom test stands: typically, device experts systematically study and characterize a sensor’s performance across a range of operating conditions.
Fourth, the characterization information is fed back to the chip designer to improve the design.
Typically, at least a few iterations of design, prototyping, and testing must be completed before the detector performance meets design expectations.
Each step in the cycle requires significant manual effort and is performed by different researchers at different institutions.
Additionally, to meet the requirements of next-generation experiments, new sensor architectures must deliver lower noise, higher sensitivity, and enhanced resolution, which necessitates more precise (and thus time-consuming) modeling and characterization.

Detector characterization infrastructure has progressed markedly, evolving from manual test benches to highly automated facilities. Early efforts concentrated on developing test systems to measure quantum efficiency, charge transfer efficiency, readout noise, and cosmetic quality under controlled conditions~\citep{Kubik_2010,DECam}. Later projects established scalable cryogenic systems with precision electronics and automated data acquisition to reproducibly process hundreds of devices~\citep{DESI_Collaboration_2022}. More recently, fully automated facilities have incorporated robotic handling, programmable illumination, and streamlined analysis pipelines~\citep{Snyder_2020,Roodman_2024}. These advances have enabled high-throughput screening of detectors, but the underlying strategy depends on sequential scanning of a predefined parameter space and expert oversight to guide testing.
As next-generation sensors introduce a wider range of tunable parameters, this approach increasingly becomes a bottleneck, placing heavy demands on laboratory testing systems.

Every aspect of the development cycle can potentially be made more efficient through automation.
Bayesian Optimization (BO) is an automation strategy that has been applied across multiple stages of sensor development.
In device design, BO has been used to optimize band gaps and materials of quantum cascade detectors via the detectivity (i.e., signal-to-noise ratio) ~\citep{poppBayesianOptimizationQuantum2021}.
In terms of operation, BO has been used in cosmic ray detection to find the optimal polarization voltage for neutron monitors in gas detectors ~\citep{GTEJEDOR20257757}.
Finally, for the operation stage of sensors, BO has been used to optimize exposure times in Electron Multiplying Charge-Coupled Devices (EMCCDs) ~\citep{bernardesOptimizationEMCCDOperating2021} and gain values for illumination invariance ~\citep{9098963}.
These studies highlight the potential of BO for streamlining sensor development, but there remains a need for integrated frameworks that connect this optimization strategy to automated test systems.

In this work, we present a tool designed to accelerate the characterization and optimization of novel sensors by integrating an automated detector testing station with machine learning (ML) optimization tools. 
We demonstrate the method on the recently-developed Single-electron Sensitive ReadOut (\sisero{}) CCD, but the framework is broadly applicable to optimizing operating parameters in other sensor architectures. 

The remainder of the paper is structured as follows.
Section \ref{sec:ccd} describes the \sisero{} CCD, amplifier design, data structures, and optimal operating conditions. 
Section~\ref{sec:opt} introduces the experimental setup for characterization, the traditional approach to sensor characterization and optimization, and the BO framework for automated optimization.
Section~\ref{sec: demo} discusses the experimental results.
Section~\ref{sec:conclusions} discusses conclusions and future prospects.

\section{\sisero{} CCD: Design and Operation}
\label{sec:ccd}


The \sisero{} CCD delivers enhanced charge sensitivity through its novel readout architecture, providing a powerful tool for faint-object astronomy, cosmology, exoplanet missions, and dark matter searches~\citep{SiSeRO_PRL_2024}.
This architecture integrates a double-gate MOSFET into the CCD sense node, which improves the precision of the charge measurement~\citep{brewer1978low}.
In addition, the sensor employs repetitive, non-destructive readout, providing sub-electron noise that enables reliable photon counting in the optical and near-infrared bands~\citep{SiSeRO_sims,SiSeRO_PRL_2024}.

The \sisero{} achieves extremely low noise and substantially faster readout compared to Skipper-CCDs \citep{Tiffenberg_2017}. 
The \sisero{} has been designed for n-channel and p-channel CCDs \citep{SiSeRO_stanford, SiSeRO_PRL_2024}. This work is focused on the \sisero{} version designed for thick, fully-depleted p-channel CCDs~\citep{SiSeRO_PRL_2024}, incorporating an isolation guard and optimized implants to support high-voltage operation while preserving sub-electron performance. 

\begin{figure}[ht!]
    \centering
    \includegraphics[width=0.5\linewidth]{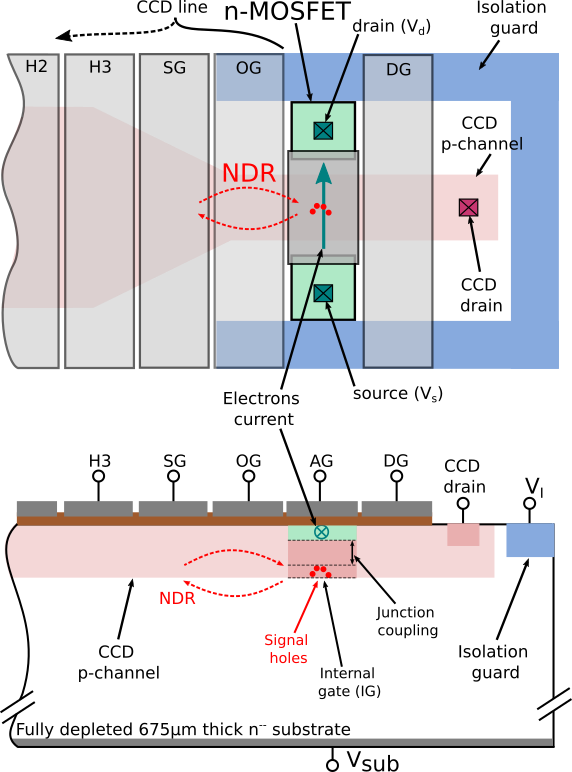}
    \caption{
    Schematic top and cross-sectional view of the \sisero{}-CCD output amplifier, consisting of an n-channel MOSFET (green shaded region) integrated into the CCD's p-channel (red shaded region). The p-channel acts as an internal gate (IG), coupling the charge (red dots) to the MOSFET. 
     During readout, charge is shifted from the CCD into the MOSFET internal gate by manipulating the voltages and timing of the clock phases of the Summing Gate (SG), Output Gate (OG) and Amplifier Gate (AG). The junction coupling between the charge packet and the transistor modulates the MOSFET's drain current, providing a high-sensitivity measurement of the pixel charge.  Non-destructive readout (NDR) is performed by shifting the charge between the SG and the \sisero~(AG), further reducing the noise on the charge measurement. The isolation guard (blue shaded region) isolates the n-type MOSFET from the n-type CCD substrate, preventing parasitic flow of electrons into the MOSFET channel.
    }
    \label{fig:nsisero}
\end{figure}

\sisero{} design studies using TCAD found that the device operates stably in depletion mode with strong junction coupling, achieving a sensitivity of $\sim$2.5\,nA per electron and a linear response for small charge packets. 
Impact-ionization noise was negligible for drain–source voltages below 5\,V. 
Noise simulations with correlated double sampling (CDS) predicted 2.4\,e$^-$\,rms/pixel at $\sim$300\,kpix/s and 0.1\,e$^-$\,rms/pixel at $\sim$700\,pix/s through multisampling---about seven times faster than Skipper-CCDs at the same noise floor~\cite{SiSeRO_sims}.
Laboratory tests validated these predictions.
The \sisero{} has record-high sensitivity with 1.54\,nA/e$^-$.
It also demonstrates sub-electron noise, reaching 0.74\,e$^-$\,rms/pixel in single-sample mode and as low as 0.021\,e$^-$\,rms/pixel with multisampling.
These results confirm single-electron resolution over a broad dynamic range, although further optimization of biasing, timing, and integration parameters will be needed to fully exploit the sensor’s potential.

The amplifier architecture of the \sisero{} sensor is shown in Fig.~\ref{fig:nsisero} ~\citep{SiSeRO_PRL_2024}. 
In order to operate the sensor, the MOSFET must be properly biased. 
The bias point is determined by the gate-to-source voltage $V_{GS}$ applied to AG and the drain-to-source voltage $V_{DS}$. Together, these voltages set the drain-to-source current $I_{DS}$, which directly controls both the sensitivity and the noise of the amplifier. 
The bias point also defines the potential well of the internal gate, thereby affecting charge transfer between the CCD and the double-gate MOSFET.
The CCD summing gate (SG) and output gate (OG) exhibit parasitic capacitive coupling to AG, introducing additional dependence of the MOSFET bias point on their voltages. 
Finally, the isolation guard voltage $V_I$ is a p-type implant that isolates the n-type region in the amplifier from the n-type substrate of the p-channel CCD array. 
It is critical in the operation of the sensor because any flow of electrons from the n-type substrate of the CCD has a direct influence on $I_{DS}$. 
On the front-end electronics, an offset voltage $V_{Offset}$ shifts the output signal from the sensor into the range of the digital readout. An improperly chosen offset voltage can push the signal outside the ADC’s linear range, leading to saturation or a loss of usable dynamic range.
Importantly, the resulting output signal reflects the combined influence of all these voltages and cannot be fully optimized by adjusting them independently.

The sensor’s output is a 2D image, where each pixel value is proportional to the accumulated charge and an added baseline from the readout electronics.
The sensor signal $S$ is the median level of the active pixels in the detector minus the baseline level.
This is estimated when the sensor is exposed to a fixed, low level of light (approximately 1000 photons per pixel).
The noise $N$ consists of amplifier noise in the sensor and other artifacts from the system readout. 
The maximum signal-to-noise ratio $\mathrm{SNR} = \frac{S}{N}$ is achieved through tuning of the bias voltages, clock levels, and clock timings. 
Traditionally, this tuning is performed by sequentially scanning a predefined parameter space and relying on expert input to identify the optimal operating conditions.

\section{Methods: Characterization and Optimization}
\label{sec:opt}

% ...
Characterizing and optimizing the operation of a CCD entails measuring the detector’s response across a range of operating parameters and analyzing the output image data to identify the optimal operating point. 
In this section, we describe the physical setup for device characterization and traditional optimization techniques.
Then we discuss a new, semi-automated characterization approach that uses Bayesian Optimization.
We apply this approach to characterize the \sisero{} CCD as a demonstration of the new method.

% ...
\begin{figure}[ht!]
    \centering
    \includegraphics[width=0.5\linewidth]{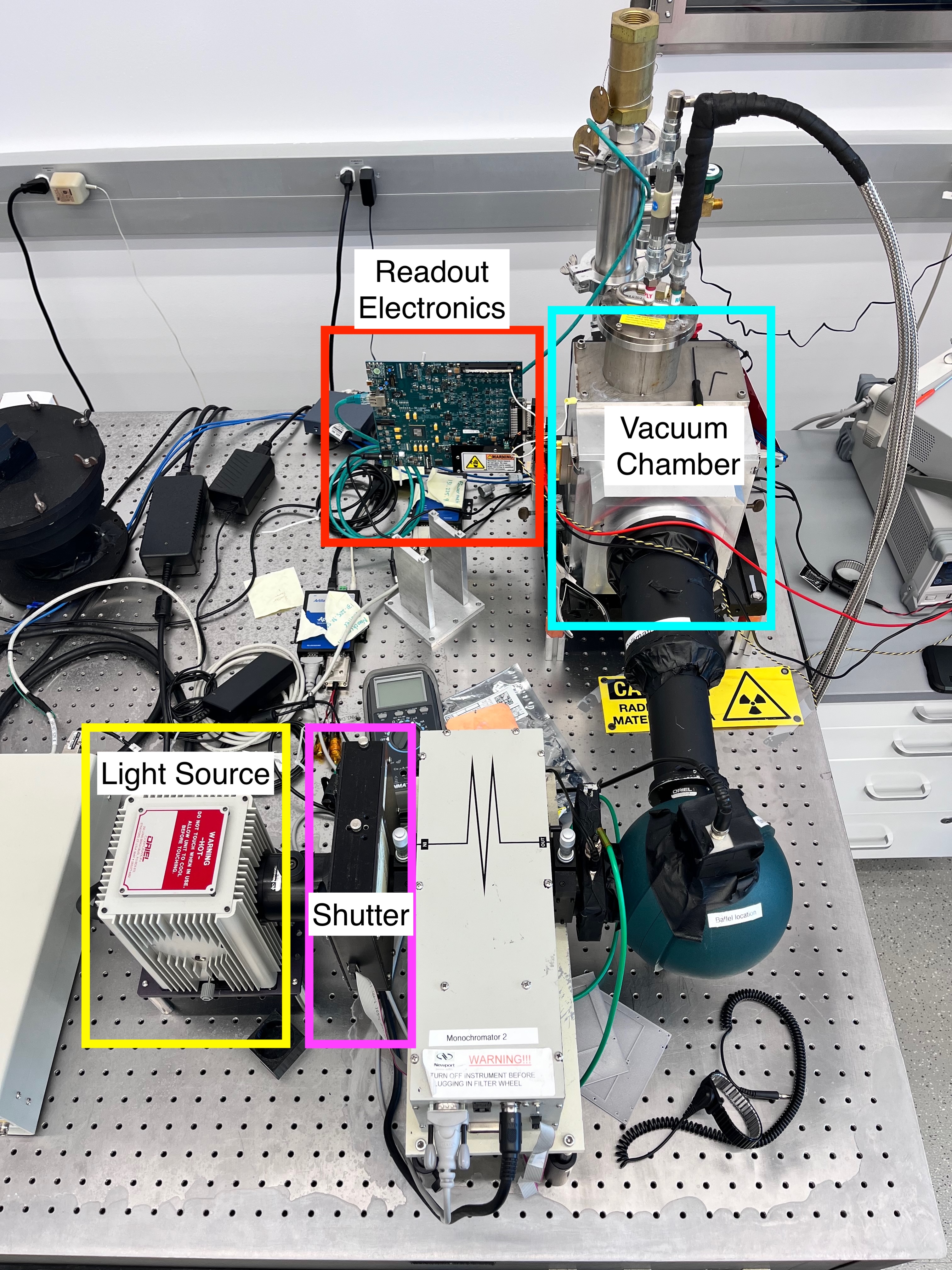}
    \caption{
    The experimental test bench setup, including the major components and the light path.
    Light is emitted from the source (yellow) and travels first to the shutter (pink), which controls the timing of illumination. 
    Light then travels to the integrating sphere which provides uniform exposure of the sensor, which resides in the vacuum chamber (cyan).
    The readout electronics (red) supply the bias voltages and clock timings to the sensor through a DB-50 port. 
    The shutter (pink) controls illumination from the light source (yellow).
    The data acquisition (DAQ) module, the black cover that is used to minimize light leaks into the system, and the cryogenics are not shown in the figure.}
    \label{fig:setup}
\end{figure}

\subsection{Experimental Setup and Characterization Procedure}

To carry out these measurements, we use a cryogenic test stand equipped with optical, mechanical, and electrical components (Figure~\ref{fig:setup}).
To stimulate the sensor's response, we illuminate it with a light source at 900\,nm, which maximizes photoelectron generation.
A mechanical shutter controls the exposure time.
The CCD is operated at $\sim$$10^{-4}$\,mbar and cooled to 130\,K to suppress noise contributions from thermal effects, such as dark current.
The power supplies and readout electronics are controlled through standard serial communication protocols.
A variety of input parameters, like bias voltages, determine the operating conditions that produce a detector response in the form of an output image.

\begin{figure*}[ht!]
    \centering
    \includegraphics[width=1\linewidth]{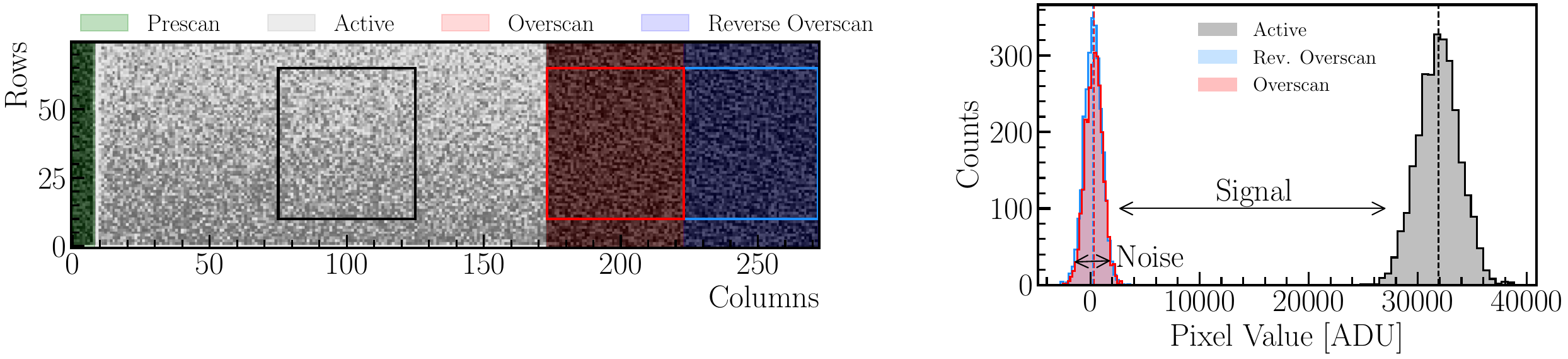}
    \caption{
    Left: An example image taken with a \sisero{}-CCD.
    The shading distinguishes regions of the image: prescan, active, overscan, and reverse overscan regions. 
    Each region is used to diagnose an aspect of the detector's performance. 
    Right: Histograms of the pixel-charge distribution corresponding to the active (black) and reverse overscan (blue) regions highlighted by open rectangles in the left panel. 
    The active pixel levels are distinct from those in the reverse overscan region, and this separation defines the signal $S$. 
    The width of the reverse overscan $\sigma_{rev}$ represents the noise in the readout system.}
    \label{fig:illustrative demonstration}
\end{figure*}

The left panel of Figure~\ref{fig:illustrative demonstration} shows an image acquired during the experiment (during optimization), which reflects the standard CCD layout: a prescan, active region, overscan, and reverse overscan.
The prescan provides a fixed electronic baseline at the start of the readout.
The active region records the stimulus-dependent signal.
The overscan (at the end of the readout) serves as a baseline diagnostic of the detector noise that includes spurious charge generated during readout.
The reverse overscan provides an additional reference of the detector noise that does not have charge-transfer artifacts.
The corresponding pixel-value distributions (right panel of  Figure~\ref{fig:illustrative demonstration}) highlight the distinct separation between signal and noise.
The values in the active region are cleanly offset from those in the overscan, and the width of the reverse-overscan distribution reflects the noise in the readout system.
Since the CCD is uniformly illuminated, such separation is indicative of an optimal image and thus provides a useful diagnostic of both sensor performance and calibration quality.

During optimization, the input operating parameters would typically be scanned sequentially under the guidance of an experienced operator, who analyzes the detector response (i.e., output image) for several metrics, including image uniformity, signal-to-noise ratio, saturation, artifacts from charge transfer, etc. 
For each round of optimization, there are four major steps: 1) set the detector operating parameters; 2) expose the sensor to light; 3) read out the image from the sensor; and 4) assess the detector response/image.
Based on their assessment, the operator adjusts the input parameters to guide the detector toward improved and more stable performance.

\subsection{Objective Function and Detector Performance Metrics}

When transitioning to automated optimization of the detector’s operating parameters, a well-defined objective function is essential to evaluate the detector's performance. During characterization, each image is analyzed to compute an objective function that rewards high signal-to-noise operating points while penalizing detector pathologies such as saturation and clock-induced charge.

We use an objective function that combines the noise-to-signal ratio with additional detector-response metrics, each evaluated for the detector operating parameters $x_i$, where $x_i$ denotes the vector of parameters used to acquire image $i$. 
The objective function is
\begin{equation}
    F(x_i)= P_0(x_i) + \sum_{j=1}^{N_p} \alpha_j P_j(x_i), 
    \label{eqn:optimization}
\end{equation}
where $P_0 = 1/SNR(x_i)$ and $P_j$ (Eqs.~\eqref{eq:penalties}-\eqref{eq:penalties2}) are detector response metrics that act as penalty functions, and $N_p$ is the number of such penalty terms. 
To compute the $SNR$, the signal $S$ is taken as the median of the active-area pixel values, while the noise $N$ is the median absolute deviation (MAD) of the reverse-overscan pixel distribution.

The detector response metrics $P_j$ are
\begin{eqnarray}
    P_1 &=& \max(0,-S)^{\frac{1}{3}}   \label{eq:penalties} \\ 
    P_2 &=& \max(0,N-S)^{\frac{1}{3}} \\ 
    P_3 &=& \max(0,2*\sigma_{rev}-\sigma_{active}) \\ 
    P_4 &=& \max(0,U-1) \\
    P_5 &=& \max(0,\mathrm{med_{ov}}-\mathrm{med_{rev}}),
    \label{eq:penalties2}
\end{eqnarray}
where the signal $S$ is the median of the active-area pixel values, and the noise estimators $\sigma_{active}$, $\sigma_{ov}$, and $\sigma_{rev}$ are the MAD of the active area, overscan, and reverse-overscan pixels, respectively (see Figure~\ref{fig:illustrative demonstration}). 
The penalty $P_1$ discourages configurations with negative gain, while $P_2$ identifies cases where the noise exceeds the signal.
$P_3$ penalizes image saturation and activates when the variance of the active-region pixels falls below the reverse-overscan noise level, indicating clipping or saturated pixels. 
$P_4$ penalizes spatial non-uniformity by evaluating an image uniformity statistic $U$ ($U\geq1)$ that divides the active region of each image into five vertical strips, computes the median signal in each strip, and compares the dispersion of these strip medians to the expected pixel-to-pixel noise level (estimated from the MAD of the full active area). 
$P_5$ estimates the level of clock-induced charge by taking the positive difference between the forward- and reverse-overscan medians -- $\mathrm{med_{ov}}$ and $\mathrm{med_{rev}}$, respectively.

The constant prefactors in Eq.~\eqref{eqn:optimization} serve as hyperparameters, selected empirically to ensure that the various penalty terms contribute comparable weight to the overall objective function. For the \sisero{} CCD optimization, the $\alpha_j$ coefficients used are 
$\alpha_1 = \tfrac{1}{2}$, 
$\alpha_2 = \tfrac{1}{2}$, 
$\alpha_3 = \tfrac{1}{3}$, 
$\alpha_4 = \tfrac{1}{250}$, 
and $\alpha_5 = 1$.
Furthermore, the five \sisero{} input parameters $x_i$ that are optimized in this work are the bias voltages and CDS integration window (see Table~\ref{tab:multiDvariables}). 
The bias voltages are strongly coupled to each other and have the greatest impact on the detector response.

% ...
\subsection{Bayesian Optimization Procedure}

\begin{figure}[htb!]
    \centering
    \includegraphics[width=0.5\linewidth]{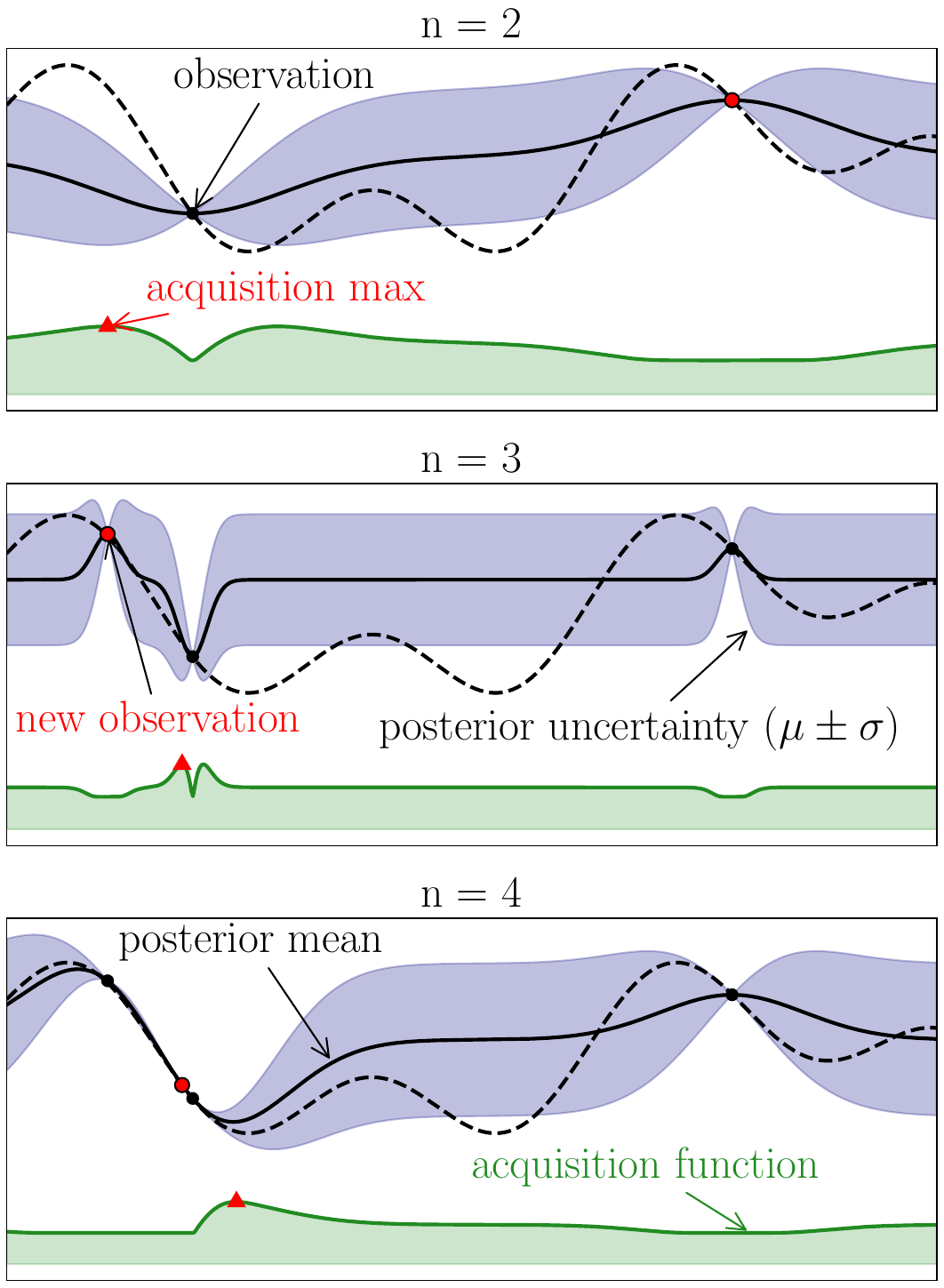}
    \caption{
    Illustration of BO showing the true objective function (dashed line) along with the evolution of the Gaussian process surrogate model's posterior mean ($\mu$; black line) and uncertainty ($\sigma$; shaded region). The surrogate model is updated over a series of $n$ observations (filled circles) where the red circle represents the $n$th observation. 
    The acquisition function (green) scores each candidate location in the parameter space, and the point at which it peaks (red triangle) is selected for the next observation.
    This illustration is inspired by Fig.~1 of~\cite{Shahriari_2016}.
    \label{fig:toy-model}
    }
\end{figure}

We use BO for automated optimization. The procedure begins by drawing an initial set of points in the five-dimensional parameter space and evaluating the objective function at those locations. BO then uses a probabilistic surrogate model of the objective function, together with an acquisition rule, to determine where to sample next. The acquisition function evaluates each candidate point according to its predicted performance and potential to reduce uncertainty in regions that have been poorly explored. At each iteration, the point that maximizes this criterion defines the next set of operating parameters, the detector response is evaluated, and the resulting objective value $F(x_i)$ is incorporated into the model. The surrogate model is updated after each observation, continually refining its estimate of the objective-function landscape. This cycle repeats for the desired number of steps, $n$.
Figure~\ref{fig:toy-model} shows an illustrative example of this process in a one-dimensional space.

% ...

\section{Optimization Experiments: Results and Discussion}
\label{sec: demo}

% ...
The experiments show that BO accelerates the optimization process when compared to the traditional procedure that relies primarily on constant expert handling.

\begin{figure*}[t!]
    \centering
    \includegraphics[width=0.9\linewidth]{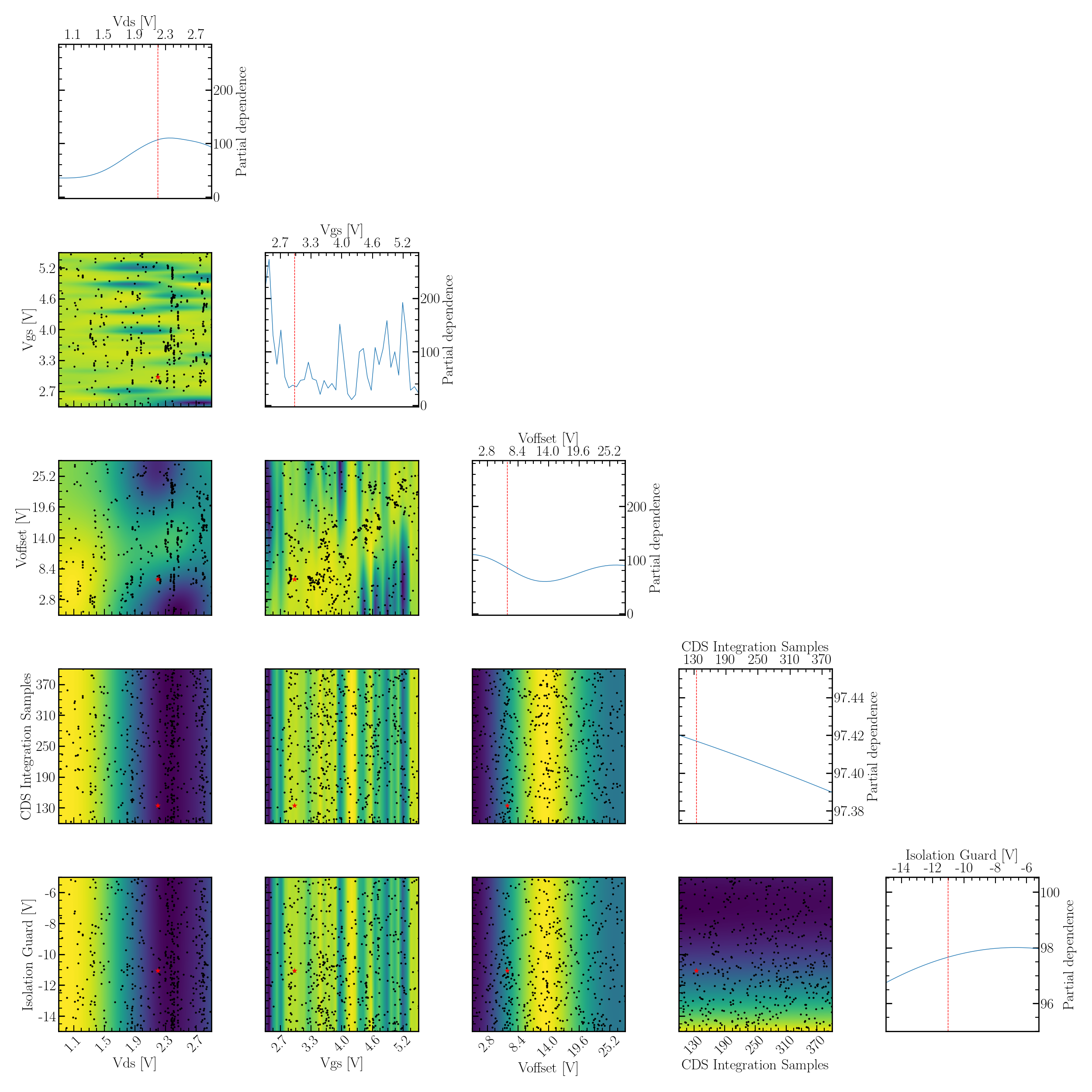}
    \caption{
    Corner plot of the objective function across the parameters used during \sisero{}-CCD optimization.
    Diagonal panels: one-dimensional partial dependence of the objective with respect to each parameter (others marginalized). 
    Off-diagonal panels: pairwise partial dependence of the objective with respect to each parameter pair. 
    The black points mark the configurations explored during burn-in and BO; dark purple indicates higher objective values, and bright yellow indicates lower values (lower objective values are better under the defined metric). 
    The red vertical markers on the diagonal panels denote the coordinate-wise minimizers (argmin) of the partial dependence curves $PD_j(\theta_j)$ over the explored range. 
    To make some weaker trends visible, the vertical scales of the CDS and $V_{I}$ marginals were expanded.
    }
    \label{fig:objective_landscape}
\end{figure*}

% ...
\begin{table}[b!]
    \centering
    \begin{tabular}{c|c|c}
    \hline
     Parameter & Description & Range \\
    \hline
    \hline
    $V_{DS}$  & Drain-to-Source Voltage & (0.9\,V, 2.9\,V) \\
    $V_{GS}$  & Gate-to-Source Voltage & (2.4\,V, 5.5\,V) \\
    $V_{Offset}$  & Offset Voltage & (0\,V, 28\,V) \\
    $V_{I}$  & Isolation Guard Voltage & ($-$6\,V, $-$15\,V) \\
    CDS  & Integration Samples & (100, 400) \\
    \hline
    \end{tabular}
    \caption{
    Parameters tuned on the detector system through the BO-instrument interface. 
    The parameters are described in Sec.~\ref{sec:ccd} and directly impact the SNR of an image.}
    \label{tab:multiDvariables}
\end{table}

% ...
We used 500 iterations---150 randomly sampled points for initialization and burn-in, followed by 350 points guided by BO.
For each iteration after burn-in, the objective function is calculated and incorporated into the surrogate model, which guides the choice of the next point.
Figure~\ref{fig:objective_landscape} displays the objective function in single dimensions (diagonal panels) and in two dimensions (off-diagonal panels) across the five-dimensional parameter space.
One-dimensional partial dependencies are the objective $F(x_i)$ versus a single parameter $\theta_k$, marginalizing the others: $$PD_k(\theta_k) = \mathbb{E}_{-\theta_k} [F(\theta_k, \theta_{-k})].$$ Across the one-dimensional partial dependencies, the BO procedure consistently identifies strong dependence on $V_{DS}, V_{GS},$ and $V_{Offset}$, while CDS integration samples and the isolation-guard voltage $V_{I}$ exhibit weaker effects.
The off-diagonal heatmaps show pairwise partial dependencies $PD_{kl}(\theta_k, \theta_l)$, with the black points showing the parameter values explored during burn-in and BO. In the pairwise partial dependence maps, the color scale represents the objective function marginalized over the remaining parameters, with higher objective values shown in dark purple and lower values in bright yellow. We note that the apparent overlap of sampled configurations with higher-objective regions in some two-dimensional projections does not imply that those configurations themselves yielded high objective values; rather, it reflects the projection of a five-dimensional optimization onto a two-dimensional subspace.

The panels in the first column of Figure~\ref{fig:objective_landscape} show the sensitivity of the objective function to $V_{DS}$ alone and when correlated with other parameters.
The objective function exhibits a broad minimum and maximum in $V_{DS}$, indicating apparent but low sensitivity to that parameter.
In the $V_{DS}-V_{GS}$ correlation, there are several small regions of maxima in the objective function, indicating relatively high sensitivity to those parameters in tandem.
This is expected because $V_{GS}$ and $V_{DS}$ set the operating point of the \sisero{} transistor, affecting the sensitivity and noise of the amplifier.
In the $V_{DS}-V_{Offset}$ correlation, there are two relatively large regions of local maxima, indicating relatively low sensitivity to those parameters in tandem.
There is a nearly monotonic gradient along the $V_{DS}$ axis with respect to CDS and $V_I$.

Next, the second column shows the degree and type of sensitivity to $V_{GS}$ alone and when correlated with three other parameters.
In the $V_{GS}-V_{offset}$ correlation, partial vertical bands exist. 
In the correlations with CDS and $V_I$, there are full vertical bands.
In the third column, the objective function with respect to $V_{Offset}$ (diagonal) shows a broad, shallow `U' shape; this is consistent with placing the baseline signal level comfortably within the ADC window while avoiding penalty regions.
In the fourth column, the CDS (diagonal) monotonically decreases because increasing the number of integration samples reduces read noise without introducing any adverse detector artifacts. 
The objective function exhibits mild curvature with respect to the isolation-guard voltage $V_{I}$.

Taken together, the partial dependence plots indicate that parameters exhibiting larger variation in their one-dimensional partial dependence, most notably the MOSFET bias voltages (\(V_{GS}\) and \(V_{DS}\)) and the offset voltage (\(V_{Offset}\)), play a dominant role in shaping the objective landscape under the present metric. The distributions of sampled configurations reflect this behavior, with non-uniform sampling patterns in the \(V_{GS}\), \(V_{DS}\), and \(V_{Offset}\) pairwise dependences, while parameters with comparatively flat one-dimensional partial dependence show no visually distinct preference. This hierarchy of parameter influence is expected to depend on both the sensor architecture and the optimization objective: for SiSeRO-class devices, the volatility of the output signal and coupling to the front-end electronics emphasize transistor biasing and ADC conversion parameters, whereas other detector architectures or optimization goals may place greater weight on CDS settings or \(V_I\).

\begin{figure}[t]

    \centering
    \includegraphics[width=0.5\linewidth]{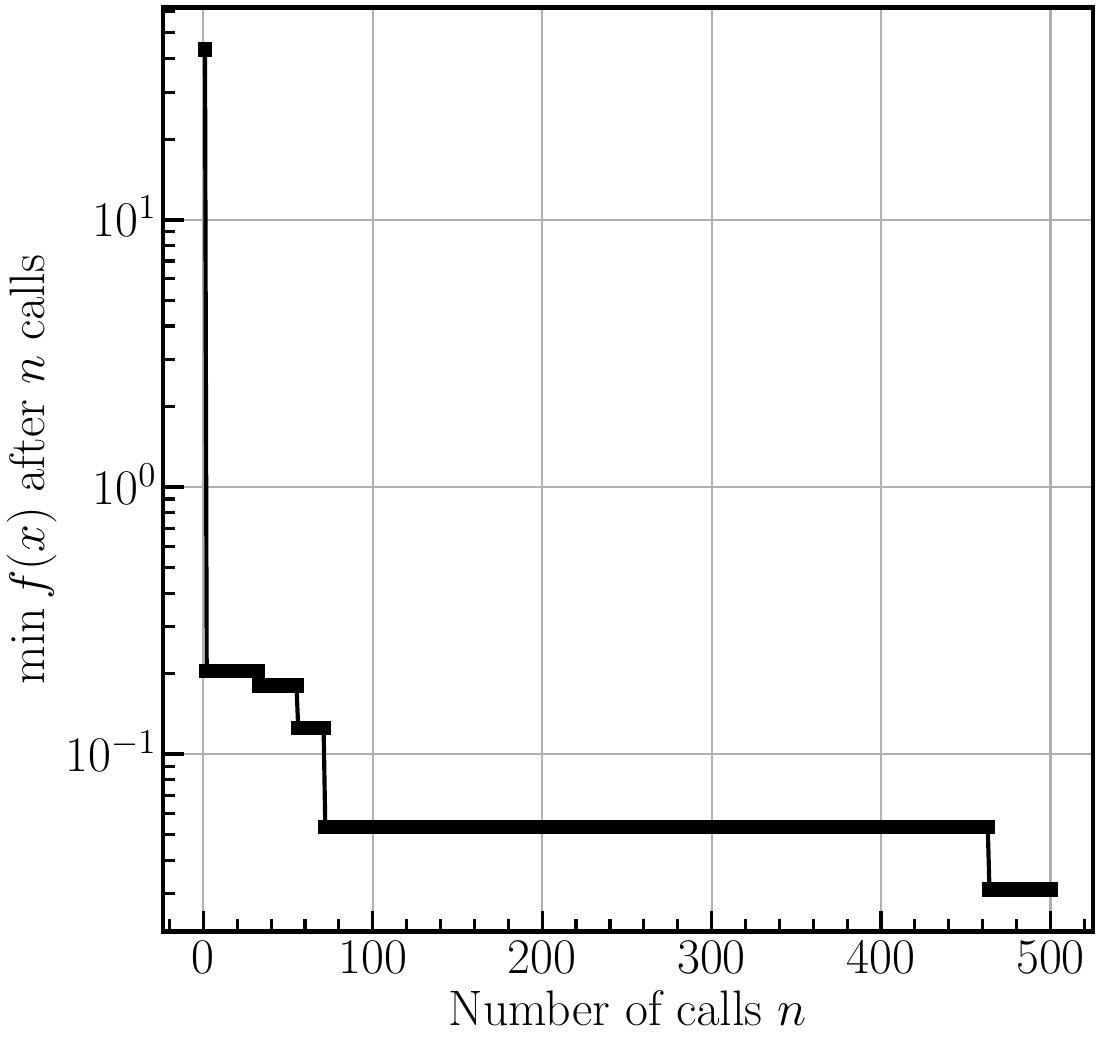}
    \caption{Running minimum of the objective versus the number of function evaluations in the 500-step study (150 random burn-in, 350 BO-guided). 
    The y-axis is on a logarithmic scale (min $f(x)$ after $n$ calls).}
    \label{fig:convergence}
\end{figure}

Figure~\ref{fig:convergence} shows the minimum of the objective value $F$ after $n$ optimization steps; this includes both the burn-in and the BO-guided phases.
The curve shows rapid improvement early on, followed by steadier gains as the optimizer refines near-favorable regions.

% ...

\section{Conclusions and Outlook}
\label{sec:conclusions}

% ...
In this work, we have demonstrated the effectiveness of a BO approach for automating the tuning of CCD sensors.
Compared to conventional manual tuning, this approach substantially reduces the time and effort required to reach near-optimal performance.

Figure~\ref{fig:convergence} illustrates the convergence behavior of the experiment discussed in Section~\ref{sec: demo}. 
This rapid convergence enabled the full tuning process to be completed in under 10 hours, whereas conventional manual scans often extend to a week or longer. 

Looking forward, the proposed technique has already been successfully validated on a Skipper CCD, confirming its robustness across different sensor architectures. 
We are extending its use to additional novel sensors~\citep{Lapi_2024} and will establish it as a standard procedure in our laboratory to streamline the characterization of new devices as they are fabricated and packaged.
Ultimately, our objective is to generalize from single-sensor optimization to the automated calibration of full sensor arrays, providing a scalable framework for next-generation detector systems.

\begin{acknowledgments}

This work was done using the resources of the Fermi National Accelerator Laboratory (Fermilab), a U.S. Department of Energy, Office of Science, Office of High Energy Physics HEP User Facility. Fermilab is managed by FermiForward Discovery Group, LLC, acting under Contract No. 89243024CSC000002. This work was supported in part by the Department of Astronomy and Astrophysics at the University of Chicago.

\end{acknowledgments}

\bibliography{main}{}
\bibliographystyle{JHEP}

\end{document}